\documentclass[twocolumn,showpacs,preprintnumbers,amsmath,amssymb,superscriptaddress,aps,prb]{revtex4}
\usepackage{graphicx}
\usepackage{dcolumn}
\usepackage{bm}
\usepackage{amsfonts}
\usepackage{mathrsfs}
\usepackage{graphicx}                   
\usepackage{dcolumn}
\usepackage{bm}
\usepackage{amsmath}

\begin{document}


\title{GdN thin-film: Chern Insulating State on Square Lattice }

\author{Zhi Li}
\affiliation{School of Materials Science and Engineering, Hefei University of Technology, Hefei 230009, Anhui, China}

\author{Jinwoong Kim}
\affiliation{Department of Physics, California State University Northridge, Northridge, California 91330-8268, USA}
\author{Nicholas Kioussis}
\affiliation{Department of Physics, California State University Northridge, Northridge, California 91330-8268, USA}

\author{Shu-Yu Ning}
\affiliation{School of Materials Science and Engineering, Hefei University of Technology, Hefei 230009, Anhui, China}

\author{Haibin Su}
\email[Email:]{hbsu@ntu.edu.sg }
\affiliation{Division of Materials Science, Nanyang Technological University, 50 Nanyang Avenue, 639798 Singapore}
\affiliation{Institute of Advanced Studies, Nanyang Technological University, 60 Nanyang View, 639673 Singapore}

\author{Toshiaki Iitaka}
\affiliation{Computational Astrophysics Laboratory, RIKEN, 2-1 Hirosawa, Wako, Saitama 351-0198, Japan}
\author{Takami Tohyama}
\affiliation{Department of Applied Physics, Tokyo University of Science, Katsushika, Tokyo 125-8585, Japan}

\author{Xinyu Yang}
\affiliation{School of Materials Science and Engineering, Hefei University of Technology, Hefei 230009, Anhui, China}

\author{Jiu-Xing Zhang}
\email[Email:]{zjiuxing@hfut.edu.cn}
\affiliation{School of Materials Science and Engineering, Hefei University of Technology, Hefei 230009, Anhui, China}

\date{\today}

\begin{abstract}
  Using first-principles calculations, we predict a Chern insulating phase in thin films of the ferromagnetic semi-metal GdN. In contrast to previously proposed Chern insulator candidates, which mostly rely on honeycomb lattices, this system affords a great chance to realize the quantum anomalous Hall Effect on a square lattice without either a magnetic substrate or transition metal doping, making synthesis easier. The band inversion between 5\emph{d}-orbitals of Gd and 2\emph{p}-orbitals of N is verified by first-principles calculation based on density functional theory, and the band gap can be as large as 100 meV within GdN trilayer. With further increase of film thickness, the band gap tends to close and the metallic bulk property becomes obvious.
\end{abstract}

\maketitle

  Topological insulator (TI), as a exotic quantum state has been experimentally realized in several materials with strong spin-orbital coupling. \cite{Kane10, Zhang11, Ando13} Chern insulator, as a material which can realize the quantum anomalous Hall effect (QAHE) also have attracted lots of research interest. The QAHE, i.e., quantum Hall effect without external magnetic field, proposed by Haldane on the hexagonal lattice,\cite{Haldane} is realized in the chromium doped (Bi,Sb)$_{2}$Te$_{3}$ thin-film recently. \cite{Fang10, Xue13, Kou14} However, the chromium-doped (Bi,Sb)$_{2}$Te$_{3}$ thin-film exhibits ferromagnetic (FM) spin order only below 15 K, and the transverse conductivity $\sigma_{xy}$ becomes quantized to better than 1/10 only below 400 mK. In addition, the QAHE also is proposed in the magnetic quantum-well (QW), \cite{CXL08,CXL13,HJ14} honeycomb materials with induced ferromagnetism (FM), \cite{Bilayer,Qiao10,Ezawa,HG,Liu13,Yao14,Yan14} magnetically doped thin-film topological crystalline insulator. \cite{Chen14} However, for the magnetic quantum well and honeycomb materials, the band gap usually is very small, and all the proposed materials are difficult to fabricate, more or less. Recently, Chern insulator with square lattice also are proposed in interface of CrO$_{2}$ and TiO$_{2}$,\cite{Cai15}, double-pervoskite monolayers \cite{ZHB14} and the magnetic rocksalt interface \cite{David13, David14}. The band gap of EuO/GdN multilayer even can be as larger as 0.1 eV,\cite{David14} in which the QAHE can persist up to room temperature.  The QW structure of Weyl semimetal (WSM), affords another simple platform to realize the QAHE.\cite{Fang11} The WSM is a type of topological matter with pairs of separated Weyl points or band crossing. The Weyl points can only appear when the spin-doublet degeneracy of each band is removed by broken time reversal symmetry or spatial inversion symmetry.\cite{Wan} For the realization of QAHE, we can consider the thin-film of WSM with broken time-reversal symmetry resulting from FM spin order. The Weyl semimetal phase in the three dimensional (3D) structure of stacked FM multilayer of TI also was proposed by Burkov et al.\cite{LB11}.

  In this work, we propose the realization of QAHE in the GdN [001] thin-film, which has a relatively large band gap for working at room temperature. Bulk GdN with rock-salt structure is known as ferromagnetic half-metal with Curie temperature T$_{c}$$\sim$58 K. \cite{60s,GdN97,GdN05,Duan07, GdN11, GdN12, Natali13} It was intensively studied in 1960s and 1990s due to its potential application in spintronics device. By first-principles calculation based on density-functional theory (DFT) and tight-binding (TB) model calculation, we predict that the band crossing in the FM GdN bulk is unavoidable even the spin orbitals coupling (SOC) is included in the first-principles calculation, and the band crossing between \emph{d}-\emph{p} orbitals in GdN bulk is protected by the four-fold rotation symmetry. The \emph{d}-\emph{p} band inversion or crossing has been discovered in several materials, such as non-centrosymmetric superconductor PbTaSe$_{2}$, \cite{PbTaSe2} non-Kondo like topological insulator YbB$_{6}$, \cite{YbB6, SmB6} topological insulator LaX (X=P, As, Sb, Bi), Weyl semimetal TaAs, \cite{WHM15} and topological semimetal LaN. \cite{LaX}  We also predict FM GdN thin-film as Chern insulator with \emph{d}-\emph{p} band inversion on square lattice.  Since the time-reversal symmetry is broken by the FM spin order of Gd 4\emph{f}-electrons, no magnetic doping or substrate is required. The \emph{d}-\emph{p} band inversion starts in the free standing GdN bilayer, and the the band gap has a maximal value $\sim$100 meV in trilayer. With the increase of thickness of thin-film, the metallic bulk character will become obvious, viz. band gap will approach to zero.


  The DFT calculations employed the all electron, full-potential linearized augment plane wave (FPLAPW) method with generalized gradient approximation plus Hubbard \emph{U} (GGA+\emph{U}) implemented in WIEN2k code. \cite{Wien2k} The SOC was included in the self-consistent calculations.  We have used the experimental value $a_{0}$=4.97 {\AA} for the lattice parameter, and the effective Hubbard $U_{eff}$ for the Gd 4\emph{f}-orbitals is set to be 9.0 eV, determined by the occupied Gd 4\emph{f} levels in GdP by optical experiment. \cite{Duan07}

\begin{figure}[t]
  \includegraphics[width=8.5cm]{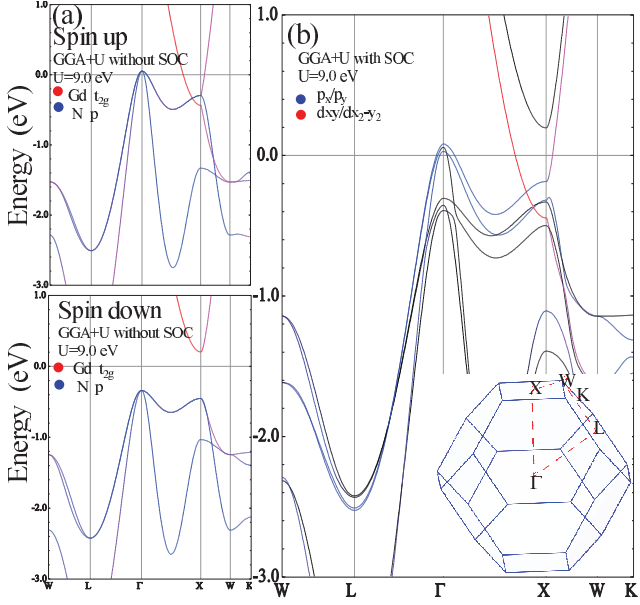}
  \caption{\label{fig1} (Color online) The band structures of FM GdN bulk by GGA+\emph{U$_{eff}$} calculation without SOC (a), and with SOC (b).In (a), the Gd $t_{2g}$-orbitals are in red, and N \emph{p}-orbitals are in blue. In (b), the Gd $d_{xy}$/$d_{x^{2}-y^{2}}$ orbitals are in red, and the N $p_{x}$/$p_{y}$ are in blue, and all the spin-down bands are in black.}
\end{figure}


  \textit{Bulk GdN} GdN is the unique FM material in the gadolinium monopnictides, and its lattice dependent Curie temperature can be up to 70 K. The electronic structure of FM bulk GdN calculated by GGA+\emph{U} without SOC is shown in Fig. 1a. With deeply buried 4\emph{f}-electron resulting from strong electronic correlation and FM Zeeman field, the Gd $d_{xy}$- and N \emph{p}-orbitals make the main contribution to the electronic structure near the Fermi level. It reveals that FM bulk GdN is half-metallic, i.e., metallic character within spin-up bands and insulating character within spin-down bands. Of interest, there is a band crossing along the $\Gamma$$\rightarrow$X direction between Gd $d_{xy}$-orbitals and N $p_{x}$($p_{y}$)-orbitals. This $d_{xy}$/ $p_{x}$($p_{y}$) band crossing is protected by the M$_{zx}$(M$_{yz}$) mirror plane, which maps y(x) to -y(-x). In the Brillouin zone (BZ)( shown in the inset of Fig. 1b), bands on the k$_{y}$=0 and k$_{x}$=0 planes can be labeled by their eigenvalues of M$_{zx}$ and M$_{yz}$, respectively. Obviously, the $p_{x}$($p_{y}$)-orbitals and $d_{xy}$-orbitals have different M$_{zx}$ (M$_{yz}$) eigenvalues 1, and -i, respectively, and they can't hybridize on the k$_{y}$(k$_{x}$)=0 plane. Along the $\Gamma$$\rightarrow$X direction, i.e.,(0,0,0)-(0,0,2$\pi$), both $p_{x}$- and $p_{y}$-orbitals can't hybridize with the $d_{xy}$-orbitals. In the TIs, SOC can remove this band crossing and open a gap, the stability of band crossing against the SOC perturbation is an important question here. The band structure of FM bulk GdN calculated by GGA+\emph{U} with SOC is shown in Fig. 1b, and it reveals that $d_{xy}$-orbitals still can't hybridize with the $p_{x}$- and $p_{y}$-orbitals to open a gap. With the spin quantization axis along z-axis, $p_{x}\pm$i$p_{y}$-orbital has quantized orbital momentum $\hbar$, and it has definite eigenvalue of $\mp$i under four-fold rotation symmetry \emph{C}$_{4z}$. However, $d_{xy}$-orbital has eigenvalue of -1 under \emph{C}$_{4z}$. \cite{note1} The band crossing between \emph{d}/\emph{p} orbitals still can be protected by the four-fold rotation symmetry \emph{C}$_{4z}$. Since the high-symmetry line $\Gamma$$\rightarrow$X is a \emph{C}$_{4z}$ invariant line, energy bands with different \emph{C}$_{4z}$ eigenvalue along this line can't hybridize with each other. So, we conclude that the \emph{d}/\emph{p} band crossing in FM bulk GdN is stable against SOC perturbation if the \emph{C}$_{4z}$ is not broken. We note that the bandwidth of Gd 5\emph{d}-orbitals is strongly dependent with the lattice constant. \cite{GdNSC,GdNopt1,GdNopt2,GdN111,Duan05} With reduced lattice parameter, both the bandwidth of Gd 5\emph{d} and the ferromagnetism will be enhanced in strained GdN. \cite{Duan05}

  The band crossing in the FM bulk GdN can be descried by a simple tight-binding model (TB) on face-centered cubic lattice with Gd \emph{d}=$d_{xy}$ and N \emph{p}=$p_{x}$+i$p_{y}$ as basis, which can be expressed as,
  \begin{eqnarray}
  \nonumber
    H &=& H_{d}+H_{p}+H_{dp}, \\   \nonumber
    H_{d} &=& \epsilon_{d}+t_{d}\sum_{i}d_{i}d_{i\pm\frac{e_{x}+e_{y}}{2}}+t_{d}'\sum_{i}d_{i}d_{i\pm\frac{e_{y}+e_{z(x)}}{2}}+h.c. \\ \nonumber
    H_{p} &=& t_{p}\sum_{i}p_{i}p_{i\pm\frac{e_{x}+e_{y}}{2}}+t_{p}'\sum_{i}d_{i}d_{i\pm\frac{e_{y}+e_{z(x)}}{2}}+h.c. \\ \nonumber
    H_{dp} &=& t_{dp}\sum_{i}p_{i}(d_{i+\frac{e_{x}}{2}}-id_{i+\frac{e_{y}}{2}}- d_{i-\frac{e_{x}}{2}}+i d_{i-\frac{e_{y}}{2}}).  \nonumber
  \end{eqnarray}

  With Fourier transformation, the Hamiltonian reads
  \begin{eqnarray}
  \nonumber
    H &=& \sum_{k}\left(
                    \begin{array}{c}
                      p_{k} \\
                      d_{k} \\
                    \end{array}
                  \right)^{\dagger}\left(
                                     \begin{array}{cc}
                                       A(k) & V(k) \\
                                       V(k)^{\dagger} & B(k) \\
                                     \end{array}
                                   \right)\left(
                                            \begin{array}{c}
                                              p_{k} \\
                                              d_{k} \\
                                            \end{array}
                                          \right)
     \\\nonumber
    A(k) &=& 4t_{p}\cos\frac{k_{x}}{2}\cos\frac{k_{y}}{2}+4t_{p}'(\cos\frac{k_{y}}{2}\cos\frac{k_{z}}{2}+
    \cos\frac{k_{x}}{2}\cos\frac{k_{z}}{2}) \\\nonumber
    B(K) &=& \epsilon_{d}+4t_{d}\cos\frac{k_{x}}{2}\cos\frac{k_{y}}{2}+4t_{d}'(\cos\frac{k_{y}}{2}\cos\frac{k_{z}}{2}+
    \cos\frac{k_{x}}{2}\cos\frac{k_{z}}{2}) \\\nonumber
    V(k) &=& 2t_{dp}(\sin\frac{k_{y}}{2}+i\sin\frac{k_{x}}{2}),
  \end{eqnarray}
  in momentum space. The A(k) and B(k)terms describe the \emph{d}- and \emph{p}-band dispersion without hybridization. The V(k) describes the hybridization on different momentum points. The vanishing \emph{d}-\emph{p} hybridization along (0,0,0)-(0,0,$\pi$)direction is obvious, and the \emph{d}-\emph{p} band crosses at $k_{z}$=0.69$\pi$, estimated by on-site energy $\epsilon_{d}$=4.2 eV and hopping parameters $t_{d}$=-0.93 eV, $t_{d}'$=0.23 eV, $t_{p}$=0.14 eV, $t_{p}'$=0.07 eV,$t_{dp}$=0.93 eV extracted from Wannier calculation with five Gd \emph{d}-orbitals and three N \emph{p}-orbitals.\cite{Wannier} With these parameter, the first Chern number is  -1 when $k_{z}$ in (-$\pi$,-0.69$\pi$) and (0.69$\pi$,$\pi$], and zero when $k_{z}$ in (-0.69$\pi$,0.69$\pi$). The calculated band structures and surface states of GdN by TB model are provided in the supplementary file. \cite{supple}

\begin{figure}[t]
  \includegraphics [width=6.5cm]{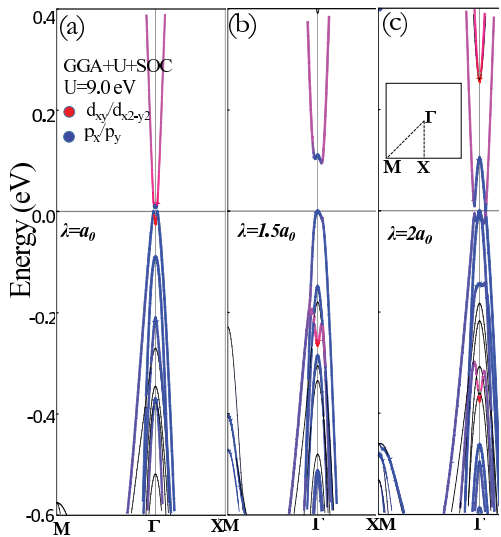}
   \caption{\label{fig2} (Color online) Band structures of GdN (AB) bilayer (a), (ABA) trilayer (b), and 2 bilayers (c), calculated by GGA+$U_{eff}$+SOC. The weight of $d_{x2-y2}$ and $p{_x}$,$p{_y}$ orbitals is represented by red and blue colors. All the spin-down bands are in black.}
\end{figure}

 \textit{Thin-film GdN}  Since the Gd $d_{xy}$ and N \emph{p} band crossing is protected by the four-fold rotation symmetry \emph{C}$_{4z}$, it's possible to remove this band crossing by size confinement, i.e., eliminate the \emph{z}-direction.  The GdN [001] thin-film, which can be regarded as a stack of GdN A and B monolayer along the \emph{z}-direction, becomes a prospective Chern insulator candidate with square lattice, instead of the extensively studied honeycomb lattice. With lattice vectors $\vec{a}=\frac{a_{0}}{\sqrt{2}}$(1,0), $\vec{b}=\frac{a_{0}}{\sqrt{2}}$(0,1), B monolayer has a offset (0.5,0.5) relative to the A monolayer. With vanishing inter-layer hopping parameter $t_{d}'$=$t_{p}'$=0 in the TB model, the $d_{xy}$ band always has higher energy than the \emph{ p}-orbitals dominating bands over the whole BZ, i.e., the monolayer of GdN should be insulating, which is consistent with our first-principles calculated result of free standing GdN monolayer. The calculated electronic structures of free standing AB bilayer (BL), ABA trilayer (TL) and 4 monolayers with thickness $\lambda$=\emph{a$_{0}$}, 1.5\emph{a$_{0}$}, and 2\emph{a$_{0}$} by first-principles calculation with GGA+$U_{eff}$+SOC are shown in Fig. 2. The spin-up Gd \emph{d}(=$d_{x2-y2}$)-orbitals in red and N \emph{p}(=$p_{x}$$\pm$ i$p_{y}$)-orbitals in blue are hybridized near the Fermi level, and all the spin-down bands are in black. Our calculated results show that the \emph{d}/\emph{p} band inversion starts in the BL, though the band gap resulting from \emph{d}-\emph{p} hybridization is minute. However, in the TL, the band gap can increase up to 100 meV, which almost is the same as the band gap in GdN/EuO interface.\cite{ David14}  Because of the enhanced bandwidth resulting from inter-layer \emph{d}-\emph{d} and \emph{p}-\emph{p} hybridization, the \emph{d}-orbitals and \emph{p}-orbitals on the B layer are inverted near the center of the BZ and hybridized and this \emph{d}/\emph{p} band inversion contributes -1 to the Chern number. In both GdN bilayer and trilayer, the A monolayer behaves like a normal 2D insulator because there is no \emph{d}-band from A layer involving with the band inversion as shown in Fig. 2b. This GdN trilayer can be regarded as Chern insulator sandwiched by ordinary insulator layer. \cite{LB11} The intra-layer \emph{d}/\emph{p} hybridization gap is $\sim$ 0.3 eV. However, the global band gap is reduced to 100 meV because of the weaker inter-layer \emph{d}/\emph{p} hybridization. With further increase of thickness, more \emph{d}/\emph{p} band inversions trend to take place. However the band gap is seriously reduced because of the vanishing long distant inter-layer \emph{d}/\emph{p} hybridization, as shown in Fig. 2c. Even though the intra-layer \emph{d}-\emph{p} hybridization can be very strong here, the \emph{p}-bands from the normal insulator layers always are obstructions to obtain large band gap. We also notice that the calculated bang gap also decreases with the increase of GdN monolayer in EuO/GdN multilayer.\cite{David14}
 \begin{figure}[t]
  \includegraphics[width=7.0cm]{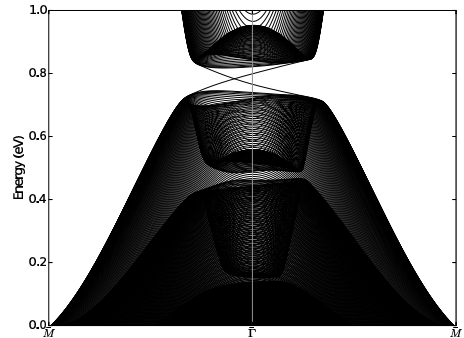}
  \caption{\label{fig3}  Calculated tight-binding electronic structure of FM GdN in form of 1D ribbon with $\epsilon_{d}$=4.2 eV, $t_{pp}$=0.14 eV, $t_{dd}$=-0.73 eV, $t'_{dd}$=0.23 eV, $t'_{pp}$=0.07 eV, $t_{dp}$=0.93 eV,$t'_{dp}$=0.05 eV. A pair of metallic surface state comes from the left and right termination of ribbon.}
\end{figure}

  The electronic structure of one dimensional ribbon of TL is calculated by the TB model with Gd $d_{x2-y2}$-orbitals and N $p_{x}$$+$ i$p_{y}$-orbiatsl as basis with on-site energy $\epsilon_{d}$=4.2 eV and $t_{d}$=0.73 eV, $t_{d}'$=0.23 eV, $t_{p}$=0.14 eV, $t_{p}'$=0.07 eV. we also include a small inter-layer \emph{d}-\emph{p} hopping parameter $t'_{dp}$=0.05 eV, which is proportional to the ($pd\pi$) integral. The calculated electronic structure of 1D ribbon is shown in Fig. 3, and two boundary states with opposite chirality, from the left and right ends of the ribbon are predicted. This metallic surface state is consistent with the prediction of quantized Hall conductivity in multilayer structure of ferromagnetic topological insulator. \cite{LB11}  We also note here that we assume the spin easy-axis is along the z-axis in all our calculation because of quantized orbital momentum along z-axis. In fact, the magnetocrystalline anisotropic energy of GdN TL is very minute and less than 2 meV, so a weak external magnetic field is necessary to keep the spin lie along z-axis at finite temperature.
 \begin{figure}[t]
  \includegraphics[width=7.0cm]{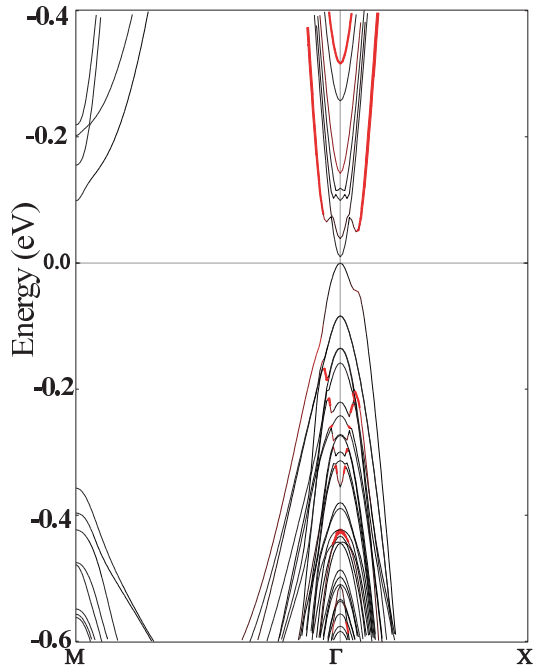}
  \caption{\label{fig4}  Calculated electronic structure of GdN TL on six YN monolayers by GGA+$U_{eff}$+SOC. The Gd $d_{x^{2}-y^{2}}$ orbitals are in red. }
\end{figure}

   For the epitaxial fabrication of [001] GdN thin-film, usually a substrate is required. Here we also calculate the electronic structure of GdN TL with six monolayers of YN. The lattice parameter of YN in rock-salt structure is 4.88 {\AA}, which is very close to the GdN lattice paramter 4.98 {\AA}. The electronic structure of this GdN/YN multilayers calculated by GGA+$U_{eff}$+SOC is shown in Fig. 4 with Gd $d_{x2-y2}$-orbitals in red, and the \emph{d}/\emph{p} band inversion is reproduced. However, The calculated band gap is about 20 meV, indicating a better substrate should be explored in future work.


  In summary, we predict the Weyl semimetal phase and Chern insulator phase in FM bulk and thin-film GdN, respectively.  In the bulk GdN, the \emph{d}-\emph{p} band crossing is protected by the four-fold rotation symmetry C$_{4z}$ , and it is robust against SOC. Chern insulating state on square lattice is predicted on the [001] thin-film GdN, and its maltilayer structure affords a concrete realization of 3D Weyl semimetal phase.  The maximal band gap is $\sim$100 meV, which presents in the free standing GdN TL with a single \emph{d}/\emph{p} band inversion. With the increasing thickness of thin-film, the band gap trends to close and metallic bulk state will become obvious.


  This is work is supported by National Natural Science Fund of China under No. 51371010 and the Fundamental Research Funds for the Central Universities. The research at CSUN was supported by NSF-PREM Grant No. DMR-1205734.  Authors also are grateful to J. P. Hu for useful discussion.



\end{document}